\documentclass[namedreferences]{SolarPhysics}
\usepackage[optionalrh]{spr-sola-addons} 
\usepackage{graphicx}        
\usepackage{color}           
\usepackage{url}             

\newcommand{\etal}{{\it et al.}}

\newcommand{\btau}{{\mbox{\boldmath$\tau$}}}
\newcommand{\bvarphi}{{\mbox{\boldmath$\varphi$}}}
\newcommand{\btheta}{{\mbox{\boldmath$\theta$}}}

\newcommand{\aap}      {{\it Astron. Astrophys.}}

\newcommand{\apj}      {{\it Astrophys. J.}}

\newcommand{\solphys}  {{\it Solar Phys.}}

\begin{document}

\begin{article}

\begin{opening}
\title{Helicity invariants of force-free field for a rectangular box \\ {\it Solar Physics}}

\author{G. V.~\surname{Rudenko}$^{1}$\sep
        I. I.~\surname{Myshyakov}$^{1}$\sep
       } \runningauthor{G.V.
Rudenko and I.I. Myshyakov} \runningtitle{Helicity invariants of
force-free field for a rectangular box}

   \institute{$^{1}$ Institute of Solar-Terrestrial Physics SB RAS, Lermontov St. 126,
Irkutsk 664033, Russia
                     email: \url{rud@iszf.irk.ru} email: \url{ivan_m@iszf.irk.ru}\\}
\begin{abstract}
An algorithm for calculating three gauge-invariant helicities
(self-, mutual- and Berger relative helicity) for a magnetic field
specified in a rectangular box is described. The algorithm is
tested on a well-known force-free model \cite {cite01} presented
in vector-potential form.

\end{abstract}
\end{opening}

\section{Introduction}
     \label{Section1}

Gauge-invariant helicities are meaningful measures of
nonpotentiality of a magnetic field in active regions. These
measures in combination with energetic estimates and their time
dependence may be used for interpretation and prediction of
various forms of solar activity (see \inlinecite {cite12};
\inlinecite {cite02}; \inlinecite {cite03}; \inlinecite {cite09}).

Self and mutual helicities correspond to the twist and writhe of
confined flux bundles, and the crossing of field lines in the
magnetic configuration respectively \cite {cite09}. Time evolution
of Berger relative helicity measures the transport of magnetic
helicity through the surface and the effect of boundary transverse
motions \cite {cite06}. These invariants represent magnetic
topology of nonpotential magnetic configurations in active regions
and its transformation caused by eruption.

This paper considers a problem of exact calculation of the
helicity invariants within a rectangular box. It is supposed that
the full magnetic field vector is prescribed everywhere in the
rectangular box. The helicity calculation problem stated in this
form as opposed to that stated for a half-space \cite {cite13} is
best suited for fixing physical characteristics of an active
region. At present, it is possible to calculate force-free spatial
magnetic field distribution in a bounded volume (see \inlinecite
{cite10}; \inlinecite {cite11}). Magnetic field extrapolation and
calculation of helicity invariants and free energy may be used for
physical analysis of solar activity. In this paper, we examine the
algorithm for calculating helicity invariants and test it on a
force-free model \cite {cite01} presented in vector-potential
form. A key feature of the algorithm is the numerical solution to
the boundary problem for a potential magnetic field ${\bf B}$ in
the rectangular box $V$ in terms of the vector potential
satisfying conditions  $\left. \nabla \cdot {\bf A}\left( {\bf
r}\right) =0\right\vert _{{\bf r}\in \overline{V}}$ and $\left.
{\bf A}\left( {\bf r}\right) \cdot \widehat{{\bf n}}\left( {\bf
r}\right) =0\right\vert _{{\bf r}\in \overline{V}}$.

\section{Basic formulations and tools description}
      \label{Section2}
\subsection{Definitions and general formulation of problems}
  \label{Section2.1}
Magnetic helicity as a gauge invariant measure may be strictly
defined only for infinite space or finite volume bounded by
magnetic surfaces. The helicity has a property to remain constant
even when magnetic reconnection dissipates energy. The term of
magnetic helicity and the condition of constant magnetic topology
can not be applied to any bounded volume because the vector
potential can not be affected by an arbitrary additional gradient
function (gauge transformation). Choosing potential field as a
reference field ${\bf B}_{ref}={\bf B}_{pot}$,
\begin{equation}\label{eq1}
\left. \left( {\bf B}_{pot}\cdot \widehat{{\bf n}}\right)
\right\vert _{\partial V}=\left. \left( {\bf B}\cdot \widehat{{\bf
n}}\right) \right\vert _{\partial V}=g,
\end{equation}
\begin{equation}\label{eq2}
{\bf B}_{pot}=\nabla \times{\bf A}_{pot},
\end{equation}
one can obtain three gauge-invariant helicity measures (see
\inlinecite {cite05}; \inlinecite {cite08} and \inlinecite
{cite06}) for the bounded volume $V$:
\begin{equation}\label{eq3}
H_{self}=\int_{V}{\bf A}_{cl}\cdot {\bf B}_{cl}dv,
\end{equation}
\begin{equation}\label{eq4}
H_{mut}=2\int_{V}{\bf A}_{pot}\cdot {\bf B}_{cl}dv,
\end{equation}
\begin{equation}\label{eq5}
\Delta H_{BF}=\int_{V}{\bf A}_{cl}\cdot \left( {\bf B}_{cl}+{\bf
B}_{pot}\right)dv- \oint_{S}\zeta \left( {\bf B}_{cl}+{\bf
B}_{pot}\right)\cdot \widehat{{\bf n}} ds.
\end{equation}
Here ${\bf B}_{cl}={\bf B}-{\bf B}_{pot}$; ${\bf A}_{cl}={\bf
A}-{\bf A}_{pot}$; $\widehat{{\bf n}}$ is the outer normal to $V$
on $S$. $\zeta$ -- the scalar potential of the gradient field
beyond $V$ whose normal component on the surface  $S=\partial V$
corresponds to that of ${\bf A}_{cl}$
\begin{equation}\label{eq6}
\left. \left( \nabla  \zeta \cdot \widehat{{\bf n}}\right)
\right\vert _{S}=\left. \left( {\bf A}_{cl}\cdot \widehat{{\bf
n}}\right) \right\vert _{S},
\end{equation}
where  ${\bf A}_{cl}$ for ${\bf B}_{cl}$  in $V$ is defined as
\begin{equation}\label{eq7}
{\bf A}_{cl}\left( {\bf r} \right)=-\frac{1}{4\pi
}\int_{V}\frac{{\bf r}-{\bf r}'}{{\left\vert { {\bf r}-{\bf
r}'}\right\vert}^{3} }\times {\bf B}_{cl}\left( {\bf r}'
\right)dv'.
\end{equation}
Using gauge invariance, one can choose the following conditions on
the vector potential
\begin{equation}\label{eq8}
\nabla \cdot {\bf A}_{pot}=0,
\end{equation}
\begin{equation}\label{eq9}
\left. \left( {\bf A}_{pot}\cdot \widehat{{\bf n}}\right)
\right\vert _{S}=0.
\end{equation}
Condition (\ref{eq9}) is useful for defining reference helicity
(\ref{eq5}). In this case, the time evolution of  $\Delta H_{BF}$
describes helicity transport through the surface (see \inlinecite
{cite06}),
\begin{equation}\label{eq10}
\frac{d}{dt }\Delta H_{BF}=-2\oint_{S}\left({\bf B}\cdot {\bf
A}_{pot}\right){\bf v}_{pot}\cdot \widehat{{\bf
n}}ds+2\oint_{S}\left({\bf v}\cdot {\bf A}_{pot}\right){\bf
B}_{pot}\cdot \widehat{{\bf n}}ds.
\end{equation}

The sum of two gauge measures (\ref{eq3}) and (\ref{eq4}) is the
reference helicity defined by \inlinecite {cite07}:
\begin{equation}\label{eq11}
\Delta H_{FA}\left( {\bf B}\right)=\int_{V}\left( {\bf A}+{\bf
A}_{pot}\right)\cdot\left( {\bf B}-{\bf
B}_{pot}\right)dv=H_{self}+H_{mut}.
\end{equation}
Reference helicities (\ref{eq5}) and (\ref{eq11}) coincide only in
special cases of $V$ - the half space or infinite space above the
sphere.

Given only a magnetic field vector in a bounded volume
$\overline{V}$, it is necessary to do the following steps to
obtain helicity invariants (\ref{eq3})-(\ref{eq5}):
\begin{description}
    \item[a)] to solve boundary problems (\ref{eq1}), (\ref{eq2}) for the vector potential ${\bf A}_{pot}$  (if ${\bf A}_{pot}$
      is known, Eq. (\ref{eq2}) gives a solution to ${\bf B}_{pot}$  for a given $g$);
    \item[b)] to obtain the vector potential of the confined magnetic field ${\bf A}_{cl}$  using Eq. (\ref{eq7});
    \item[c)] to obtain the scalar potential $\zeta$ on $\partial V$  corresponding to (\ref{eq6}).
\end{description}
In this paper, we solve problems a) and c) for the rectangular
domain: $V=\left(0, L_{x}\right)\times \left(0, L_{y}\right)\times
\left(0, L_{z}\right)$. As mentioned above, problem b) has already
its analytical solution (\ref{eq7}).
\subsection{Problem a)}
  \label{Section2.2}
  Let
\begin{equation}\label{eq12}
g\in L^2(S).
\end{equation}
Consider the problem of finding the vector potential ${\bf A}^0
\in C^{\infty }(\overline{V})$ satisfying equations:
\begin{equation}\label{eq13}
\nabla \times \nabla \times {\bf A}^0={\bf 0},\qquad\nabla \cdot
{\bf A}^0=0 \qquad in \qquad\overline{V}.
\end{equation}
and the boundary condition
\begin{equation}\label{eq14}
{\left\Vert (\nabla\times{\bf A}^0) \cdot\widehat{{\bf
n}}-g\right\Vert}_{L^2(S)}=0 .
\end{equation}
If it is possible to solve the problem (\ref{eq12})-(\ref{eq14}),
the vector-potential ${\bf A}_{pot}$ equation corresponding to
(\ref{eq9}) can be written as
\begin{equation}\label{eq15}
{\bf A}_{pot}={\bf A}^0-\nabla\times{\bf A}^1,
\end{equation}
where  ${\bf A}^1$ is the solution to (\ref{eq13}) with boundary
conditions
\begin{equation}\label{eq16}
{\left\Vert (\nabla\times{\bf A}^1) \cdot\widehat{{\bf
n}}-\left.{\left( {\bf A}^0\cdot\widehat{{\bf
n}}\right)}\right\vert_S\right\Vert}_{L^2(S)}=0 .
\end{equation}
The solution to (\ref{eq12})-(\ref{eq14}) can be written as a sum:
\begin{equation}\label{eq17}
{\bf A}^0=\sum^6_{i=1}{\bf A}^{g_i}+{\bf A}^m.
\end{equation}
Here ${\bf A}^{g_i}$ is the solution to (\ref{eq13}) with the
boundary function in (\ref{eq14})
\begin{equation}\label{eq18}
g_i({\bf r})=\left(%
\begin{array}{cc}
  g({\bf r})-G_i; & {\bf r} \in S_i \\
  0; & {\bf r} \in S_j, j\neq  i \\
\end{array}%
\right),
\end{equation}
\begin{equation}\label{eq19}
G_i=\int_{S_i}gds=\int_{S_i}\nabla\times {\bf A}^m
\cdot\widehat{{\bf n}}_ids,
\end{equation}
where $S_i$  - the $i$-th side of the rectangular box $V$;
 ${\bf A}^m$  is the superposition of vector potentials given by the system of 5 magnetic monopoles located beyond the box
 $V$
\begin{equation}\label{eq20}
{\bf A}^m=\sum^5_{i=1}\frac{m_i\widehat{{\bf z}}_i\times({\bf
r}-{\bf r}_i)}{|{\bf r}-{\bf r}_i|\left[|{\bf r}-{\bf r}_i|+({\bf
r}-{\bf r}_i)\cdot \widehat{{\bf z}}_i\right]}.
\end{equation}
This system of monopoles satisfying Equations (\ref{eq19})
 can be easily constructed by fixing their ${\bf
r}_i$ locations and $\widehat{{\bf z}}_i$ orientations in space
and solution of linear algebraic system of equations for unknown
magnitudes $m_i$ (orts $\widehat{{\bf z}}_i$ should be chosen so
that tails of monopoles do not cross the volume $V$). To each
problem (\ref{eq13}),(\ref{eq18}) on ${\bf A}^{g_i}$ there is a
solution giving a unique field $\nabla \times{\bf A}^{g_i}$ , and
thus there is a solution to (\ref{eq12})-(\ref{eq14}) on ${\bf
A}_0$, giving a unique ${\bf B}_{pot}$.

Due to the freedom to choose gradient normalization, we can
reformulate problem (\ref{eq13}),(\ref{eq18}) on ${\bf A}^{g_i}$
in terms given by \inlinecite {cite04}:
\begin{equation}\label{eq21}
\Delta {\bf A}^{g_i}={\bf 0} \quad in \quad \overline{V},
\end{equation}
\begin{equation}\label{eq22}
\nabla \cdot {\bf A}^{g_i}=0 \quad in \quad \overline{V},
\end{equation}
\begin{equation}\label{eq23}
\nabla_t \cdot {\bf A}^{g_i}_t=0 \quad on \quad S,
\end{equation}
\begin{equation}\label{eq24}
{\bf A}^{g_i}_t=\nabla ^{\perp }\chi=\nabla_t \chi\times
\widehat{{\bf n}}_i \quad on \quad S_i,
\end{equation}
\begin{equation}\label{eq25}
\partial_n {\bf A}^{g_i}_n=0  \quad on \quad S,
\end{equation}
\begin{equation}\label{eq26}
\widehat{{\bf n}}_i \cdot {\bf A}^{g_i}=0 \quad in \quad
\overline{V},
\end{equation}
\begin{equation}\label{eq27}
-\nabla ^2_t\chi=g_i \quad on \quad S_i,
\end{equation}
\begin{equation}\label{eq28}
\partial_n \chi=0  \quad on \quad \partial S_i.
\end{equation}
Here the subscript $t$ stands for the trace of the operator or the
field on the boundary; the derivative $\partial_n \chi$  in
Equation (\ref{eq28}) means the derivative of the normal to the
surface $\partial S_i$ in the plane of $S_i$.
  \subsubsection{BVP for the vector-potential function given on one side of the rectangular box}
\label{Section2.2.1} Further, without loss of generality, is
assumed that index $i$ corresponds to the side lying in the plane
$(z=0)$. Then $A^{g_i}_z=0$  from Equation (\ref{eq26}). If $\chi$
is known, tangential components $A_x$ and $A_y$ are BVP solutions
to (\ref{eq21})-(\ref{eq26}) and satisfy the properties for
$\partial S_i$, which follow from Equations (\ref{eq24}) and
(\ref{eq28}):
\begin{equation}\label{eq29}
\begin{array}{c}
  A^{g_i}_x(x,0,0)=\partial_y \chi(x,0,0)=0, \quad A^{g_i}_x(x,L_y,0)=\partial_y \chi(x,L_y,0)=0, \\
  \partial_x A^{g_i}_x(0,y,0)=\partial_x \partial_y \chi(0,y,0)=0, \\
  \partial_x A^{g_i}_x(L_x,y,0)=\partial_x\partial_y \chi(L_x,y,0)=0; \\
\end{array}
\end{equation}
\begin{equation}\label{eq30}
\begin{array}{c}
  A^{g_i}_y(0,y,0)=-\partial_x \chi(0,y,0)=0, \quad A^{g_i}_y(L_x,y,0)=-\partial_x \chi(L_x,y,0)=0, \\
  \partial_y A^{g_i}_y(x,0,0)=-\partial_x \partial_y \chi(x,0,0)=0, \\
  \partial_y A^{g_i}_y(x,L_y,0)=-\partial_x\partial_y \chi(x,L_y,0)=0. \\
\end{array}
\end{equation}
Properties (\ref{eq29}) and (\ref{eq30})) allow us to choose
suitable orthonormal bases for $A^{g_i}_x$  and $A^{g_i}_y$ on
$S_i$  in $L^2$
\begin{equation}\label{eq31}
\begin{array}{c}
h_x^{mn}=\sqrt{\frac{2-\delta
_{m0}}{L_{x}}}\sqrt{\frac{2}{L_{y}}}\cos(\pi mx/L_x)\sin(\pi
ny/L_y), \\
m={ 0,1,...,\infty}, \quad  n={ 1,2,...,\infty} ;\\
\end{array}
\end{equation}
\begin{equation}\label{eq32}
\begin{array}{c}
h_y^{mn}=\sqrt{\frac{2-\delta
_{n0}}{L_{y}}}\sqrt{\frac{2}{L_{x}}}\sin(\pi mx/L_x)\cos(\pi
ny/L_y), \\
m={ 1,2,...,\infty}, \quad  n={ 0,1,...,\infty} .\\
\end{array}
\end{equation}
Here $L_x$, $L_y$  are the scales of $S_i$. Basis functions
(\ref{eq31}) and (\ref{eq32}) satisfy (\ref{eq29}) and
(\ref{eq30}), therefore expansions $A^{g_i}_x$ and $A^{g_i}_y$ in
these bases converge according to the norm of $W^1$ in the
neighborhood of $\partial S_i$ (i.e., their expansions converge to
them smoothly on $\partial S_i$).

Using Equations (\ref{eq31}) and (\ref{eq32}), one can find a
harmonical solution in $\overline{V}$ to problem
(\ref{eq21})-(\ref{eq28}):
\begin{equation}\label{eq33}
{\bf A}^{g_i}=\left(%
\begin{array}{c}
  \sum^{\infty}_{m=0} \sum^{\infty}_{n=1} a_x^{mn} h_x^{mn}(x,y)p^{mn}(z) \\
  \\
  \sum^{\infty}_{m=1} \sum^{\infty}_{n=1} a_y^{mn} h_y^{mn}(x,y)p^{mn}(z) \\
  \\
  0 \\
\end{array}%
\right).
\end{equation}
Here
\begin{equation}\label{eq34}
    p^{mn}(z)=\frac{e^{-q^{mn}z}-e^{-q^{mn}(2L_z-z)}}{1-e^{-q^{mn}L_z}},
    \quad q^{mn}=\sqrt{(\pi m /L_x)^2+(\pi n /L_y)^2},
\end{equation}
\begin{equation}\label{eq35}
    a^{mn}_x=\int_0^{L_x}\int_0^{L_y}\partial_y
    \chi(x,y)h^{mn}_x(x,y)dxdy
\end{equation}
\begin{equation}\label{eq36}
    a^{mn}_y=-\int_0^{L_x}\int_0^{L_y}\partial_x
    \chi(x,y)h^{mn}_y(x,y)dxdy
\end{equation}
Solution (\ref{eq33}) satisfies (\ref{eq24}) in $L^2$ and strictly
satisfies (\ref{eq21})-(\ref{eq23}), (\ref{eq25}) and
(\ref{eq26}). Choosing suitable bases (\ref{eq31}) and
(\ref{eq32}) provides smooth solutions in the neighborhood of
$\partial S_i$ . The magnetic components expressed through partial
derivatives ${\bf A}^{g_i}$ can be easily represented by
analytical expressions of (\ref{eq33}). The same method is applied
to all sides of $V$. In numerical implementation of this scheme,
expansions (\ref{eq33}) may be limited by the number of terms
corresponding to dimensions of the grid $\chi(x,y)$ . In our
implementation, we show integrals of (\ref{eq35}) as a sum of
analytical integrals in grid cells, representing functions
$\partial_y \chi$  and $\partial_x \chi$ as a 2D linear
interpolation in the cell of their grid values.
 \subsubsection{BVP-$\chi$}
\label{Section2.2.2} Let us consider the same side $(z=0)$ as in
previous item. The solution to (\ref{eq27}), (\ref{eq28}) will be
presented as follows:
\begin{equation}\label{eq37}
    \chi({\bf r})=\chi^*-\chi^0-\sum^4_{j=1}\chi^{\tau_j},
\end{equation}
where
\begin{equation}\label{eq38}
    \chi^*=\frac{1}{2\pi}\int_{S_i}g_i({\bf
    r}')\ln\frac{1}{\left\vert {{\bf r}-{\bf r}'}\right\vert }dx'dy'
\end{equation}
is the inhomogeneous solution to (\ref{eq27}). The second and
third components are potential functions satisfying the
two-dimensional homogeneous Laplace equation. $\chi^0$
  will be defined as superposition of the potentials
\begin{equation}\label{eq39}
    \begin{array}{c}
      \chi^0=m_1x+m_2y+m_3xy+m_4(x^2-y^2)+m_5(3yx^2-y^3)+ \\
      m_6(3xy^2-x^3)+m_7(x^3y-xy^3)+m_8(6x^2y^2-x^4-y^4)+ \\
      m_9(5xy^4-10x^3y^2+x^5)+m_{10}(5yx^4-10x^2y^3+y^5)+ \\
      m_{11}(y^6-x^6-15y^4x^2+15x^4y^2),\\
    \end{array}
\end{equation}
where the vector of coefficients $m$ is selected to satisfy eight
conditions of equality between tangential derivatives $\chi^0$ and
$\chi^*$ at four vertices of $S_i$  and three conditions of
equality between $\chi^0$ and $\chi^*$ integrals on any three
edges of  $S_i$. Thus, given that the surface integral of $g_i$ is
$0$, the difference of $\chi^*-\chi^0$ have zero tangential
derivatives at vertcies of $S_i$ and zero means on all edges of
$S_i$. Let us label $S_i$ edges by $j$ and define boundary
univariate functions $\tau_j$ equal to values of the normal
derivative of $\chi^*-\chi^0$ on corresponding edges $j$.
$\chi^{\tau_j}$ will be defined as potential functions satisfying
the Neumann problem and the boundary condition:
\begin{equation}\label{eq40}
    \begin{array}{c}
      {\left\Vert {\partial_n \chi^{\tau_k}-\tau_k}\right\Vert}_{L^2}=0 \quad on \quad (\partial S_i)_k, \quad k=j \quad and  \\
      \partial_n =0 \quad on \quad (\partial S_i)_k, \quad k\neq j.\\
    \end{array}
\end{equation}
Without loss of generality, let $j$ corresponds to an edge on the
line $y=0$. The solution to the problem $\chi^{\tau_j}$  in this
case can be written in the following form:
\begin{equation}\label{eq41}
    \begin{array}{c}
     \chi^{\tau_j}=-\sum^{\infty
    }_{m=1}b^m\sqrt{\frac{2}{L_x}}\cos(\pi
    mx/L_x)\frac{1}{q^m}\left({\frac{e^{-q^m(2L_y-y)}+e^{-q^my}}{1-e^{-q^mL_y}}}\right),\\
    q^m=\frac{\pi m}{L_x},\\
  \end{array}
\end{equation}
\begin{equation}\label{eq42}
    b^m=\int^{L_x}_0\tau_j(x,y)\sqrt{\frac{2}{L_x}}\cos(\pi
    mx/L_x)dx.
\end{equation}

Solution (\ref{eq41}) satisfies (\ref{eq40}) in $L^2$  and
strictly satisfies (\ref{eq27}), (\ref{eq28}) in $\overline{S}_i$.
Solution (\ref{eq41}) is smooth at the vertices of $S_i$.
Components $A^{g_i}_x$ and $A^{g_i}_y$ expressed through partial
derivatives $\chi^{\tau_j}$ can be described by the analytical
expressions of (\ref{eq41}). The same method is applied to every
edge of $S_i$. In numerical implementation of this scheme,
expansions (\ref{eq41}) can be limited by the number of terms
corresponding to grid dimensions $\tau_k(x,y)$. In our
implementation, we show integrals of (\ref{eq42}) as a sum of
analytical integrals in grid cells, representing functions
$\tau_k$ as a 2D linear interpolation in the cell of its grid
values.
\subsection{Problem c)}
  \label{Section2.3}
  To obtain potential $\zeta$  satisfying Equation (\ref{eq6}) on the surface $S$, we use (\ref{eq7}). Equation (\ref{eq7})
   yields ${\bf A}_{cl}$  on the surface $S'$  limiting the volume
   $V' \subset  V$
\begin{equation}\label{eq43}
    V'=(-\Delta x,L_x+\Delta x)\times(-\Delta y,L_y+\Delta y)\times(-\Delta z,L_z+\Delta
    z),
\end{equation}
where  $\Delta x,\Delta y,\Delta z$, are scales of the elementary
cell of the discrete grid defined on $V$. To calculate $\zeta$, we
first estimate its values on the grid $S'$ . To do this, we
specify its zero value at an arbitrary vertex of $V'$
$\zeta_0({\bf r'}_0)=0$ and then calculate integrals on the broken
curves $\Gamma_i$ jointing vertices ${\bf r'}_i$ on $S'$ and ${\bf
r'}_0$:
\begin{equation}\label{eq44}
    \zeta'({\bf r'}_i)=\int^{{\bf r'}_i}_{{\bf r'}_0}{\bf
    A}_{cl}\cdot{\bf \btau}_idl,
\end{equation}
where  $\btau$ is the unit local tangential vector to $\Gamma_i$.
To calculate $\zeta$ on $S$, we add integrals
\begin{equation}\label{eq45}
    \zeta({\bf r}_i)=\zeta'({\bf r'}_i)-\int^{{\bf r}_i}_{{\bf r}_0}{\bf
    A}_{cl}\cdot\widehat{\bf n}dl.
\end{equation}
Integrals (\ref{eq44}), (\ref{eq45}) are approximated by simple
summation. Integrals (\ref{eq44}) do not depend on $\Gamma_i$
curve shape. Therefore in their discrete implementation the choice
of different configurations of $\Gamma_i$ should result in small
differences between results of integration of (\ref{eq44}),
(\ref{eq45}).
\section{Results of model calculations of helicity invariants}
      \label{Section3}
Let us take the calculation of helicity invariants
(\ref{eq3})-(\ref{eq5}), (\ref{eq11}) as an example for the
analytical force-free model \cite {cite01}. The expression for the
magnetic field will be written in terms of vector-potential for
$n=1$:
\begin{equation}\label{eq46}
    {\bf A}_{LL}=-\frac{1}{r^2}W\widehat{\bf
    r}+U\frac{1}{r\sqrt{1-\mu^2}}\widehat{\bvarphi },
\end{equation}
\begin{equation}\label{eq47}
    W=-a\int^{\mu}_0\frac{P^2(\mu')}{\sqrt{1-\mu'^2}}d\mu',
\end{equation}
where $\mu=\cos \theta$,
\begin{equation}\label{eq48}
    U=\frac{P(\mu)}{r},
\end{equation}
$P(\mu)$ satisfies nonlinear differential equation
\begin{equation}\label{eq49}
    (1-\mu^2)\frac{d^2P}{d\mu^2}+2P+2a^2P^3=0
\end{equation}
with boundary values
\begin{equation}\label{eq50}
    P(-1)=P(1)=0.
\end{equation}
Applying operator $\nabla\times$  to (\ref{eq46}) yields
\begin{equation}\label{eq51}
    {\bf B}_{LL}=\nabla\times {\bf
    A}_{LL}=\frac{1}{r\sqrt{1-\mu^2}}\left(\frac{1}{r}\partial_\theta U \widehat{\bf
    r}-\partial_rU\widehat{\btheta }+aU^2\widehat{\bvarphi } \right)
\end{equation}
that coincides with Equation (\ref{eq3}) from \inlinecite
{cite01}.

For the tests we employed the force-free model calculated in the
rectangular box $V$  ($L_x=L_y=1$, $L_z=0.8$; $N_x=N_y=100$,
$N_z=80$) with the following parameters: $a^2=a^2_{1,1}=0.425$;
${\bf r}_c=[0.5,0.5,-0.25]$ are coordinates of the source; the
axis of dipole lies in the plane $(x,z)$ and has an angle
$\Phi=45^o$ to the $z$-axis.

Next, the notations in (\ref{eq3})-(\ref{eq5}) and (\ref{eq11})
for helicity invariants are used to denote their calculated
values. To estimate the accuracy of the algorithm, we have also
calculated helicity invariant (\ref{eq11}) in terms of the known
vector potential ${\bf A}_{LL}$:
\begin{equation}\label{eq52}
\Delta H_{FA,LL}\left( {\bf B}\right)=\int_{V}\left( {\bf
A}_{LL}+{\bf A}_{pot}\right)\cdot\left( {\bf B}-{\bf
B}_{pot}\right)dv.
\end{equation}
Relative helicity (\ref{eq5}) was calculated for two different
sets  $\{ \Gamma_i \}$ (\ref{eq44}) and marked with indices
$\{1,2\}$. All helicity invariants were normalized to the true
(not invariant!) helicity
\begin{equation}\label{eq53}
H_{LL}=\int_{V}{\bf A}_{LL}\cdot {\bf B}_{LL}dv.
\end{equation}

Table 1 lists the results. Numbers in Table 1 are helicities of
the selected model characteristics. The relative errors indicating
the level of completion of gauge invariance of Finn \& Antonsen
reference helicity (3 and 4 columns) and the degree of accuracy of
Berger reference helicity (5 and 6 columns) are $0.0025$ and
$0.0134$ respectively. Such sufficiently small errors justify the
application of the algorithm presented here to calculate helicity
invariants of real active regions.

\begin{table}
  \centering
  \caption{ }\label{t1}
\begin{tabular}{cccccc}
  \hline
  $H_{self}/H_{LL}$ & $H_{mut}/H_{LL}$ & $\Delta H_{FA}/H_{LL}$ & $\Delta H_{FA,LL}/H_{LL}$ & $\Delta H_{BF,1}/H_{LL}$ & $\Delta H_{BF,2}/H_{LL}$ \\
  \hline
  0.1166 & 0.6009 & 0.7175 & 0.7157 & 0.7916 & 0.7810 \\
  \hline
\end{tabular}
\end{table}

\end{article}


\begin{thebibliography}{}
\bibitem[\protect\citeauthoryear{{Amari} \etal}{2003a}]{cite02}
Amari,~T., Luciani,~J.~F., Aly,~J.~J., Mikic,~Z., and Linker,~J.:
2003a, \apj{} \textbf{585}, 1073.

\bibitem[\protect\citeauthoryear{{Amari} \etal}{2003b}]{cite03}
Amari,~T., Luciani,~J.~F., Aly,~J.~J., Mikic,~Z., and Linker,~J.:
2003a, \apj{} \textbf{595}, 1231.

\bibitem[\protect\citeauthoryear{{Amari} \etal}{1999}]{cite04}
Amari,~T., Boulmezaoud,~T.~Z., and Mikic,~Z.: 1999, \aap{}
\textbf{350}, 1051.

\bibitem[\protect\citeauthoryear{{Berger}}{1999}]{cite05}
Berger,~M.~A.: 1999, {\it in Magnetic Helicity in Space and
Laboratory Plasmas}, ed. M. R. Brown, R. C. Canfield, and A. A.
Pevtsov, 1

\bibitem[\protect\citeauthoryear{{Berger} and {Field}}{1984}]{cite06}
Berger,~M.~A., and Field,~G.~B.: 1984, {\it J. Fluid Mech.}
\textbf{147}, 133.

\bibitem[\protect\citeauthoryear{{Brown} and {Priest }}{1999}]{cite12}
Brown,~D.~S., and Priest,~E.~R.: 1999, \solphys{} \textbf{190},
25.

\bibitem[\protect\citeauthoryear{{Finn} and {Antonsen}}{1985}]{cite07}
Finn,~J.~M., and Antonsen,~T.~M.: 1985, {\it Comments Plasma Phys.
Controlled Fusion} \textbf{9}, 111.

\bibitem[\protect\citeauthoryear{{Longcope} and {Malanushenko}}{2008}]{cite13}
Longcope,~D.~W., Malanushenko,~A: 2008, \apj{} \textbf{674}, 1130.

\bibitem[\protect\citeauthoryear{{Low} and {Lou}}{1990}]{cite01}
Low,~B.~C., Lou,~Y.~Q.: 1990, \apj{} \textbf{352}, 343.

\bibitem[\protect\citeauthoryear{{Priest}}{1999}]{cite08}
Priest,~E.~R.: 1999, {\it in Magnetic Helicity in Space and
Laboratory Plasmas}, ed. M. R. Brown, R. C. Canfield, and A. A.
Pevtsov, 141

\bibitem[\protect\citeauthoryear{{Regnier} \etal}{2005}]{cite09}
Regnier,~S., Amari,~T., and Canfield,~R.~C.: 2005, \aap{}
\textbf{442}, 345.

\bibitem[\protect\citeauthoryear{{Rudenko} and {Myshyakov}}{2009}]{cite10}
Rudenko,~G.~V., and Myshyakov,~I.~I.: 2009, \solphys{}
\textbf{257}, 287.

\bibitem[\protect\citeauthoryear{{Rudenko} \etal}{2010}]{cite11}
Rudenko,~G.~V., Myshyakov,~I.~I., and Anfinogentov,~S.~A.: 2011,
eprint arXiv:1007.0298, (\solphys{}, in press)





\end{thebibliography}
\end{document}